\documentclass[11pt]{article}
\usepackage{geometry}
\usepackage{graphicx,times}
\usepackage{amssymb,bm,colortbl,natbib,booktabs,amsmath}

\newcommand{\alphadot}{\ensuremath{\alpha_\bullet}}
\newcommand{\alphadota}{\ensuremath{\alpha_{\bullet {a{+}1}}}}
\newcommand{\dd}{\ensuremath{\,\text{d}}}
\newcommand{\mdm}{\ensuremath{\text{\sl MDM}}}

\newcommand{\mc}[3]{\multicolumn{#1}{#2}{#3}}
\newcommand{\lc}[1]{\mc{1}{l}{\cellcolor{white}\ensuremath{#1}}}
\newcommand{\cc}[1]{\mc{1}{c}{\cellcolor{white}\ensuremath{#1}}}

\newcommand{\br}{\bm{r}}
\newcommand{\bn}{\bm{n}}
\newcommand{\bnsd}{\bn_{*\bullet}}
\newcommand{\bnds}{\bn_{\bullet *}}
\newcommand{\ba}{\bm{\alpha}}

\newcommand{\sumn}{\sum_{\bn\in\mathcal{N}}}
\newcommand{\nid}{n_{i\bullet}}
\newcommand{\nda}{n_{\bullet a}}
\newcommand{\rid}{r_{i\bullet}}
\newcommand{\rda}{r_{\bullet a}}
\newcommand{\ndd}{n_{\bullet\bullet}}
\newcommand{\rdd}{r_{\bullet\bullet}}
\newcommand{\aldd}{\alpha_{\bullet}}
\newcommand{\ala}{\alpha_a}

\newcommand{\E}{\ensuremath{\mathbb{E}}}
\newcommand{\var}{\ensuremath{\mathbb{V}ar}}
\newcommand{\cov}{\ensuremath{\mathbb{C}ov}}

\usepackage[affil-it]{authblk}

\usepackage[sc]{mathpazo}
\usepackage[T1]{fontenc}
\providecommand{\keywords}[1]{\textbf{Keywords:} #1}

\begin{document}





\title{The multivariate Dirichlet-multinomial distribution \\%
  and its application in forensic genetics to adjust for
  sub-population effects using the $\theta$-correction}

\author{T.~TVEDEBRINK\thanks{Corresponding author} ~~~ P.~S.~ERIKSEN}
\affil{Department of Mathematical Sciences, Aalborg University,
  \\Fredrik Bajers Vej 7G, DK-9220 Aalborg East,
  Denmark\\ { \normalsize \texttt{tvede@math.aau.dk} ~~ \texttt{svante@math.aau.dk}}}
 
\author{ N.~MORLING}
\affil{Section of Forensic Genetics, 
  Department of Forensic Medicine, \\Faculty of Health and Medical Sciences,
  University of Copenhagen, \\Frederik V's Vej 11, DK-2100
  Copenhagen, Denmark\\ {\normalsize \texttt{niels.morling@sund.ku.dk}}}


\maketitle

\begin{abstract}
  In this paper, we discuss the construction of a multivariate
  generalisation of the Dirichlet-multi\-nomial distribution. An example
  from forensic genetics in the statistical analysis of DNA mixtures
  motivates the study of this multivariate extension.

  In forensic genetics, adjustment of the match probabilities due to
  remote ancestry in the population is often done using the so-called
  $\theta$-correction. This correction increases the probability of
  observing multiple copies of rare alleles and thereby reduces the
  weight of the evidence for rare genotypes.

  By numerical examples, we show how the $\theta$-correction
  incorporated by the use of the multivariate Dirichlet-multinomial
  distribution affects the weight of evidence. Furthermore, we
  demonstrate how the $\theta$-correction can be incorporated in a
  Markov structure needed to make efficient computations in a Bayesian
  network.
  \vskip2mm
  \noindent\keywords{Multivariate Dirichlet-multinomial distribution; STR DNA mixture;
  Forensic genetics; $\theta$-correction}
\end{abstract}


\section{Introduction}
\label{sec:intro}

When biological material is obtained from a scene of crime, it is
often possible to produce a DNA profile from even minute amounts of
DNA. In cases where DNA from more than one individual is present in
the resulting DNA profile, the DNA profile is called a DNA
mixture. DNA mixtures are harder to interpret and analyse than single
contributor stains as there are many sources of uncertainty, e.g.\ the
number of contributors, the relative amounts of contributed DNA and
the individual DNA profiles of the contributors. For more than twenty
years \citep{evett1991}, statistical modelling of DNA mixtures has
attracted much attention. The statistical models have been extended to
cope with more of the uncertainties and artifacts observed in the
detected mixed DNA profile. Modelling these components is important in
order to assess the probability of the evidence, since it is the task
of the forensic geneticists to assign an evidential weight by
computing the likelihoods of the evidence under competing hypotheses.

Recently, \citet{cowell2014} published a statistical model for DNA
mixtures, which in a coherent framework enabled the modelling of
common phenomena as stutters (artefacts of the polymerase chain
reaction, PCR), allelic drop-out (undetected alleles of the true
contributors) and silent alleles (unobservable alleles, e.g.\ due to
mutations in primer binding regions). In order to estimate the model
parameters, the authors maximised the likelihood under each
hypothesis. Due to the vast number of possible combinations of DNA
profiles, this is computationally demanding and challenging. However,
the methodology of \citet{cowell2014} and its implementation
\citep[R-package \texttt{DNAmixtures},][]{dnamixtures2014} solved this
by utilising Bayesian networks and the implementation of these in the
Hugin Software (\texttt{http://www.hugin.com}).

As future work, \citet[][Section~5.3.2]{cowell2014} suggested to
implement a correction for subpopulation effects on the allele
probabilities.  In order to correct for these subpopulation structures,
\citet{nichols1991} suggested the ``$\theta$-correction'' to be used
when inferring the weight of evidence in forensic genetics.  The
Markov structure for representing the individual genotypes in
\citet{cowell2014} imposed to conform with the Bayesian network
paradigm did not allow for incorporating correlation between the
individual DNA profiles \cite[see Fig.~\ref{fig:markov} below and also
Fig.~4 of][]{cowell2014}.

Here, we show how this Markov structure can be modified in order to
incorporate positive correlations between alleles within and among the
genotypes involved. A consequence of positive correlation is an
increased probability of homozygosity, which may be induced by
subpopulation structures in the
population. 

The resulting distribution when incorporating the $\theta$-correction
for multiple contributors in a Bayesian network framework is a
multivariate generalisation of the Dirichlet-multinomial distribution.
The Dirichlet-multinomial distribution was first derived by
\citet{mosimann1962}, who derived it as a compound distribution in
which the probability vector of a multinomial distribution is assumed
to follow a Dirichlet distribution \citep{mosimann1962}. After
marginalisation over this distribution, the cell counts follow a
Dirichlet-multinomial distribution \citep{mosimann1962,johnson1997}.

The present paper is structured as follows: In
Section~\ref{sec:dirmult}, we discuss how the $\theta$-correction is
implemented for a single DNA profile. In Section~\ref{sec:more}, this
is generalised for more contributors. This multiple contributor
extension of the genotype model leads to the introduction of the
multivariate Dirichlet-multinomial distribution. In
Section~\ref{sec:mdm}, we derive the structures of the marginal and
conditional distributions of the multivariate Dirichlet-multinomial
distribution. Furthermore, the expression of the generalised factorial
moments is derived, which is used to obtain the mean and covariance
matrix of the distribution. In Section~\ref{sec:results}, we show by
numerical examples how the $\theta$-correction affects the weight of
the evidence.

\section{Dirichlet-multinomial distribution}
\label{sec:dirmult}

In order to adjust for genetic subpopulation structures when
computing the weight of evidence in forensic genetics, it is common to
use the $\theta$-correction \citep{balding1994}. Several authors have
discussed the interpretation of $\theta$; \citet{weir1999,curran2002}
derived likelihood ratio expressions with $\theta$ being the
probability that a pair of alleles is identical-by-descent
(IBD). \citet{tvedebrink2010} defined $\theta$ as an overdispersion
parameter in a multinomial sampling scheme, and \citet{green2009}
discussed $\theta$ in relation to assumptions made about founding
genes in populations.

\begin{figure}[!h]
  \centering
  {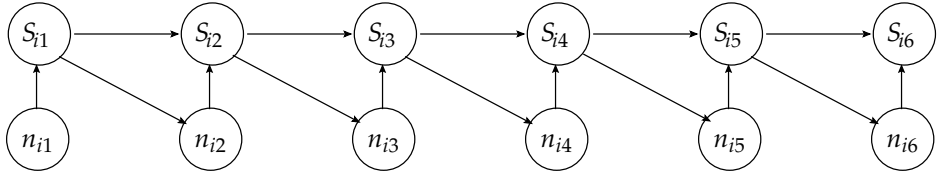}
  \caption{\label{fig:markov}Markov structure of allele counts of
    contributor $i$ for a marker with six possible alleles.}
\end{figure}

In forensic genetics, the prevailing genotyping technology is based on
short tandem repeat (STR) loci. The genotype at a given STR locus is
represented by a pair of alleles, each of which is inherited from the
individual's parents. Let $A$ denote the possible number of alleles,
typically in the range of five to 20, at a given STR locus. The
genotype of individual $i$ can be represented as a vector of allele
counts, $\bn_i$, where $n_{ia}$ is the number of $a$ alleles of the
genotype. By forming a cumulative sum, $S_{ia} = \sum_{b=1}^a n_{ia}$,
of allele counts, $n_{ia}$, for alleles $a\in\{1,\dots,A\}$,
\citet{graversen2014} showed that the multinomial distribution over
allele counts for unknown contributors may be evaluated by the product
of a sequence of binomial distributions (see Fig.~\ref{fig:markov})
such that $n_{ia}\mid S_{i,a-1} \sim \text{bin}(2-S_{i,a-1},Q_a)$,
where $Q_a=q_a/\sum_{b=a}^A q_b$.

If the distribution of allele probabilities is assumed to follow a
Dirichlet distribution, then the marginal distribution of allele
counts under a multinomial sampling scheme follows a
Dirichlet-multinomial distribution \citep{tvedebrink2010}. Using
similar derivations as in \citet{graversen2014}, we show that the
$\theta$-correction can be incorporated by evaluating the
Dirichlet-multinomial distribution by a sequence of beta-binomial
distributions.

Let $n=\sum_{b=1}^A n_b$ and suppress the subscript $i$, then the
Dirichlet-multinomial distribution can be specified by
\begin{displaymath}
  P(n_{1},\dots,n_{A}) =
  \frac{n!\Gamma(\alphadot)}{\Gamma(n+\alphadot)}\prod_{b=1}^A
  \left\{\frac{\Gamma(n_{b}+\alpha_b)}{n_{b}!\Gamma(\alpha_b)}\right\},
\end{displaymath}
where $\alphadot = \sum_{b=1}^A\alpha_b$ and
$\bm{\alpha}=(\alpha_1,\dots,\alpha_A)$ being positive real valued
parameters \citep{johnson1997}. The joint distribution over sums of
disjoint subsets of cell counts is also Dirichlet-multinomial
\citep[][pp.~81]{johnson1997}. In particular, when collapsing the last
$A-a$ cells into one cell it will yield a parameter-vector of
$(\alpha_1,\dots,\alpha_{a},\alphadota)$ with
$\alphadota=\sum_{b=a{+}1}^A\alpha_b$. In the case where $\bn$ denotes
the allele counts, $n=2$ and the distribution of allele counts
$(n_1,\dots,n_a)$, $a\in\{1,\dots,A{-}1\}$ is given by:
\begin{displaymath}
  P(n_1,\dots,n_a) =
  \frac{2!\Gamma(\alphadot)}{\Gamma(2{+}\alphadot)}
  \frac{\Gamma(2{-}S_a{+}\alphadota)}{(2{-}S_a)!\Gamma(\alphadota)}
  \prod_{b=1}^a
  \left\{\frac{\Gamma(n_b{+}\alpha_b)}{n_b!\Gamma(\alpha_b)}\right\}.
\end{displaymath}
Using this result, we obtain the conditional distribution of $n_a$
given $n_1,\dots,n_{a-1}$ as
\begin{align*}
  P(n_a\,|\, n_{a{-}1}\dots,n_1) &= 
  \frac{P(n_a, n_{a{-}1}\dots,n_1)}{P(n_{a{-}1}\dots,n_1)}\\[2mm]
  &= \dfrac{\dfrac{2!\Gamma(\alphadot)}{\Gamma(2+\alphadot)}
  \dfrac{\Gamma(2-S_a+\alphadota)}{(2-S_a)!\Gamma(\alphadota)}
  \prod\limits_{b=1}^a
  \left\{\dfrac{\Gamma(n_b+\alpha_b)}{n_b!\Gamma(\alpha_b)}\right\}}%
{\dfrac{2!\Gamma(\alphadot)}{\Gamma(2+\alphadot)}
  \dfrac{\Gamma(2-S_{a{-}1}+\alpha_{\bullet{a}})}{(2-S_{a{-}1})!\Gamma(\alpha_{\bullet{a}})}
  \prod\limits_{b=1}^{a-1}
  \left\{\dfrac{\Gamma(n_b+\alpha_b)}{n_b!\Gamma(\alpha_b)}\right\}}\\[2mm]
  &= {2{-}S_{a{-}1}\choose n_a} 
  \frac{\Gamma(\alphadota+\alpha_a)}{\Gamma(\alpha_a)\Gamma(\alphadota)}
  \frac{\Gamma(n_a{+}\alpha_a)\Gamma(2{-}S_{a{-}1}{-}n_a+\alphadota)}{\Gamma(2{-}S_{a{-}1}+\alphadota+\alpha_a)},
\end{align*}
where from the second to the third line, we used that
$S_a=S_{a{-}1}+n_a$ and $\alphadota = \alpha_{\bullet{a}} -
\alpha_a$. This is a beta-binomial distribution
\citep[][pp.~81]{johnson1997} with parameters
$(2-S_{a-1},\alpha_a,\alphadota)$ that are similar to those of the
binomial distribution $(2-S_{a-1},Q_a)$. Similarly to
\citet{graversen2014}, we observe directly from the expression that
\mbox{$n_{a}\perp\!\!\!\perp (n_1,\dots,n_{a-1},S_1,\dots,S_{a-2})\mid
  S_{a-1}$}.  Finally, we note that the allele probabilities
$\bm{q}=(q_1,\dots,q_A)$ and $\theta$ are related to $\bm{\alpha}$
through $q_a = \alpha_a/\alphadot$ and $\theta=(1+\alphadot)^{-1}$
\citep{tvedebrink2010}.

\section{Multivariate Dirichlet-multinomial distribution}
\label{sec:more}

\begin{figure}[!h]
  \centering 
  {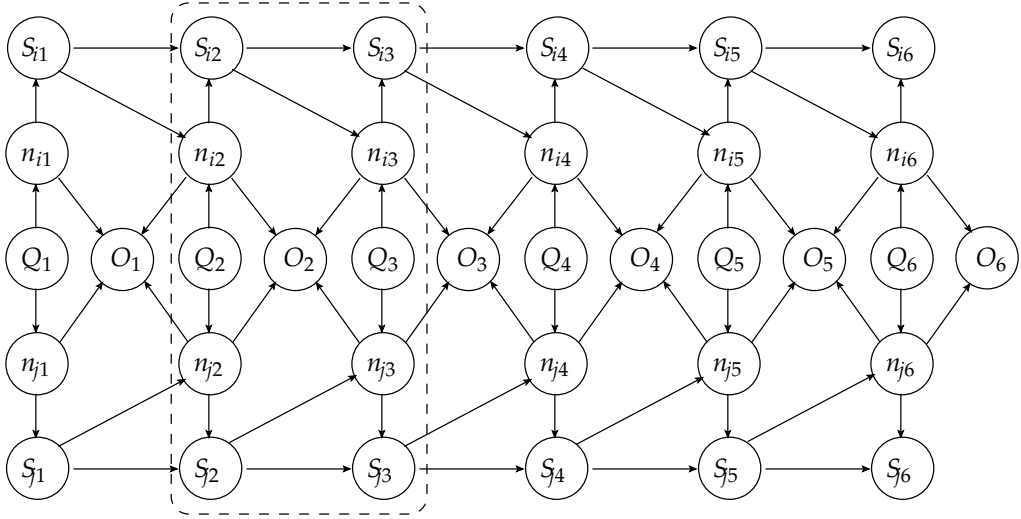}
  \caption{\label{fig:dirmult}Extended Markov structure for
    incorporating $\theta$-correction for two contributors. As in
    Fig.~\ref{fig:markov}, there are six possible alleles, where $Q_a$
    denotes the scaled allele probabilities. For profile $i$, the
    allele counts and cumulative sums are given by $(n_{ia},S_{ia})$,
    and similarly $(n_{ja},S_{ja})$ for profile $j$. The $O_a$ nodes
    have the same meaning as in \citet{cowell2014}. The dashed box
    represents the clique necessary to propagate in the Bayesian
    network.}
\end{figure}

When incorporating the $\theta$-correction for more contributors, it
is necessary to modify the Markov structure in Fig.~\ref{fig:markov}
as we need to model the joint distribution of $n_{ia}$ and $n_{ja}$ in
order to incorporate the positive correlation from remote ancestry.
Thus, the Markov structure depicted in \citet[][Fig.~4]{cowell2014}
should be replaced by the Markov structure in Fig.~\ref{fig:dirmult}.

In Fig.~\ref{fig:dirmult}, the distribution of the allele
probabilities, $q_a$, was modelled by a Dirichlet distribution. This
distribution can be specified sequentially by the following relation:
$Q_{a} \sim \text{beta}(\alpha_a,\alphadota)$, where
$Q_a=q_a/\sum_{b=a}^A$ for $a=1,\dots,A{-}1$. Furthermore, these
beta-distributions are mutually independent \citep{johnson1997}, which
implies that the Dirichlet distribution can be formulated as a product
of beta distributions.

First, we observe that, conditioned on $Q_a$ and cumulative sums, the
allele counts from the two individuals are mutually independent:
\begin{align*}
  P(n_{ia},n_{ja}\mid S_{i,a{-}1},S_{j,a{-}1},Q_a) &= P(n_{ia}\mid
  S_{i,a{-}1},Q_a)P(n_{ja}\mid S_{j,a{-}1},Q_a)\\
  &= {2{-}S_{i,a{-}1}\choose n_{ia}}{2{-}S_{j,a{-}1}\choose n_{ja}}
  Q_a^{n_{\bullet a}}(1-Q_a)^{4-S_{\bullet a{-}1}-n_{\bullet a}},
\end{align*}
where $n_{\bullet a}=n_{ia}+n_{ja}$ and $S_{\bullet
  a{-}1}=S_{i,a{-}1}+S_{j,a{-}1}$. Secondly, we marginalise over $Q_a$,
which is beta distributed with parameters $(\alpha_a,\alphadota)$:
\begin{multline*}
  \int_0^1 P(n_{ia}\mid S_{i,a{-}1},Q_a)P(n_{ja}\mid
  S_{j,a{-}1},Q_a)f(Q_a)\dd Q_a\\=
  {2{-}S_{i,a{-}1}\choose n_{ia}}{2{-}S_{j,a{-}1}\choose n_{ja}}
  \frac{\Gamma({\alphadot}_a)}{\Gamma(\alpha_a)\Gamma(\alphadota)}
  \int_{0}^1
  Q_a^{n_{\bullet a}+\alpha_a-1} (1-Q_a)^{4-S_{\bullet
      a{-}1}-n_{\bullet a}+\alphadota-1} \dd Q_a, 
\end{multline*}
which is the integral of a non-normalised beta-distribution. By
letting $S_{i0} = S_{j0} = 0$, we have for $1\le a<A$ that
$P(n_{ia},n_{ja}\mid S_{i,a{-}1},S_{j,a{-}1})$ is given by:
\begin{equation}
  {2{-}S_{i,a{-}1}\choose n_{ia}}{2{-}S_{j,a{-}1}\choose n_{ja}}
  \frac{\Gamma({\alphadot}_a)}{\Gamma(\alpha_a)\Gamma(\alphadota)}
  \frac{\Gamma(n_{\bullet a}+\alpha_a)\Gamma(4-S_{\bullet a{-}1}-n_{\bullet a}+\alphadota)}%
  {\Gamma(4-S_{\bullet a{-}1}+{\alphadot}_a)}.
  \label{eq:tvedebrink}
\end{equation}

A consequence of marginalising over $Q_a$ is that the clique size in
the network decreases. For the two profiles in Fig.~\ref{fig:dirmult},
this marginalisation implies that the relevant clique size decreases
from eight to six nodes as $Q_{a{-}1}$ and $Q_a$ are removed, while
the imposed correlation connects the nodes $n_{ia}$ and $n_{ja}$
(graph not shown).

In the general setting, where we consider a DNA mixture of $I$
contributors, we denote $\bm{n}=(\bm{n}_{1},\dots,\bm{n}_{I})$, where
each $\bm{n}_i = (n_{i1},\dots,n_{iA})$ denotes the allele counts for
profile $i$ and, similarly, for the cumulative sums, $S_{ia} =
\sum_{b=1}^a n_{ib}$. Hence, $P(n_{1a},\dots,n_{Ia}\mid
S_{1,a{-}1},\dots,S_{I,a{-}1})$ is given by
\begin{displaymath}
  \left\{\prod_{i=1}^I {2{-}S_{i,a{-}1}\choose n_{ia}}\right\}
  \frac{\Gamma({\alphadot}_a)}{\Gamma(\alpha_a)\Gamma(\alphadota)}
  \frac{\Gamma(n_{\bullet a}+\alpha_a)\Gamma(2I - S_{\bullet a{-}1} -
    n_{\bullet a}+\alphadota)}{\Gamma(2I-S_{\bullet
      a{-}1}+{\alphadot}_a)},
\end{displaymath}
where $n_{\bullet a}=\sum_{i=1}^I n_{ia}$ and $S_{\bullet
  a{-}1}=\sum_{i=1}^I S_{i,a{-}1}$.

In full generality, consider a set of vectors
$\bn=(\bn_1,\dots,\bn_I)$, where $\bn_i=(n_{i1},\dots,n_{iA})$ and
$n_{i\bullet} = \sum_{a=1}^A n_{ia}$ for
$n_{i\bullet}\in\mathbb{Z}_0$. Then the probability mass function for
$\bn$ is given by
\begin{equation}
  P(\bn) = \left\{\prod_{i=1}^I
    {n_{i\bullet}\choose\bn_i}\right\}
  \frac{\Gamma(\aldd)}{\Gamma(\ndd+\aldd)} 
  \prod_{a=1}^A \frac{\Gamma(n_{\bullet a} +
    \alpha_a)}{\Gamma(\ala)},
  \label{eq:multdirmult}
\end{equation}
where $\ndd = \sum_{a=1}^A \nda = \sum_{a=1}^A\sum_{i=1}^I
n_{ia}$. For a single contributor, i.e.\ $I=1$ and
$\bn=(n_1,\dots,n_A)$, this distribution simplifies to the
Dirichlet-multinomial distribution. Hence, we may call this
distribution the multivariate Dirichlet-multinomial (MDM)
distribution, which we denote $\mdm(\bnsd,\ba)$, where
$\bnsd=(n_{1\bullet},\dots,n_{i\bullet},\dots,n_{I\bullet})$ is the
vector of trails per experiment (row sums in
Table~\ref{tab:contingency}) or e.g.\ the number of alleles per DNA
profile. Furthermore, we observe from \eqref{eq:multdirmult} that
inference about the model parameters, $\bm{\alpha}$, only depends on
$\bnds=(n_{\bullet1},\dots,n_{\bullet a},\dots,n_{\bullet A})$, i.e.\
the column sums shown in Table~\ref{tab:contingency}.

\begin{table}
  \caption{\label{tab:contingency}Sufficient statistics of a table
    when modelled using the multivariate Dirichlet-multinomial (MDM)
    distribution. By construction of the MDM distribution, the
    row sums, $\nid$, and the total sum, $\ndd$, are known and fixed.} 
  \centering
  \begin{tabular}{>{$}l<{$}*{6}{>{$}c<{$}}}
    & 1 & \dots & a & \dots & A &\\   
    \rowcolor[gray]{0.8} \lc{1} & n_{11} & \dots & n_{1a} & \dots &  n_{1A} & \cc{n_{1\bullet}}\\
    \rowcolor[gray]{0.8} \lc{\vdots} & \vdots & \ddots & \vdots & \ddots & \vdots & \cc{\vdots} \\
    \rowcolor[gray]{0.8} \lc{i} & n_{i1} & \dots & n_{ia} & \dots & n_{iA} & \cc{n_{i\bullet}}\\
    \rowcolor[gray]{0.8} \lc{\vdots} & \vdots & \ddots & \vdots & \ddots & \vdots & \cc{\vdots} \\
    \rowcolor[gray]{0.8} \lc{I} & n_{I1} & \dots & n_{Ia} & \dots & n_{IA} & \cc{n_{I\bullet}}\\[2mm]
    \lc{}& \lc{n_{\bullet 1}} & \lc{\dots} & \lc{n_{\bullet a}} & \lc{\dots} & \lc{n_{\bullet A}} & \lc{n_{\bullet \bullet}}\\
  \end{tabular}
\end{table}

To emphasise the difference between row and column marginals, we let
\linebreak
$\bnds=(n_{\bullet1},\dots,n_{\bullet a},\dots,n_{\bullet
  A})$ denote the column sums, which we shall use in the next section
when discussing conditional and marginal distributions.

Furthermore, let $B\subset\{1,\dots,A\}$ be a subset of the cells,
e.g.\ a subset of the alleles in a genetics context, and let $C$
denote the complement of $B$. The counts associated with $B$, $C$ and
the row sums over $C$ are defined by
\begin{multline*}
  \bn_{*B} = \{n_{ia}\}_{a\in B},~~ \bn_{*C} = \{n_{ia}\}_{a\in C}
  ~~\text{and}~~
  \bnsd^{(C)} = \left\{\sum\nolimits_{a\in C} n_{ia}\right\} 
  ~~\text{for}~~ i=1,\dots,I,\text{~respectively.}
\end{multline*}
Similarly, we may consider subsetting over index $i$ such that $J$
and $K$ specify two disjoint and exhaustive partitions of
$\{1,\dots,i,\dots,I\}$, where $\bn_{J*}$ and $\bn_{K*}$ denote the
counts, respectively. In the DNA mixture context, this corresponds to
partition of the set of $I$ contributors into two disjoint groups.

\section{Properties of multivariate Dirichlet-multinomial
  distribution}
\label{sec:mdm}

\subsection{Conditional and marginal distributions}
\label{sec:cond-marg}

The construction of the MDM distribution implies that it carries many
similarities to the Dirichlet-multinomial distribution. For the MDM
distribution, one may consider marginalisation and conditioning over
both $i$ and $a$ in the $n_{ia}$ notation. Furthermore, we may also
condition on $\bnds$ and $\bnsd$ to obtain a generalisation of the
hypergeometric distribution. 

First, we consider the marginal and conditional distribution over
index $a$: The marginal distribution of $\bn_{*B}$ can be thought of
as the distribution when collapsing all elements of $C$ into one
hyper-class (or allele). By using similar arguments as
\citet[][pp.~81]{johnson1997}, this gives the results that the
marginal and conditional distributions are MDM with parameters given
by
\begin{displaymath}
 \bn_{*B}\sim\mdm(\bnsd,\{\ba_B,\alpha_C\}) 
 \quad\text{and}\quad
 \bn_{*B}\mid\bn_{*C}\sim\mdm(\bnsd-\bnsd^{(C)},\ba_B),
\end{displaymath}
where $\ba_B=\{\alpha_a\}_{a\in B}$ and $\alpha_C = \sum_{a\in C}
\alpha_a$.

Secondly, we handle the case of marginalising and conditioning over
index $i$. It follows directly from \eqref{eq:multdirmult} that the
distribution of $\bn_{J*}$ is MDM with parameters $\bn_{J\bullet}$ and
$\ba$, where $\bn_{J\bullet} = \{n_{i\bullet}\}_{i\in J}$. The
conditional distribution of $\bn_{J*}$ given $\bn_{K*}$ can be
considered as a posterior distribution as we have already observed
counts $\bn_{K*}$, which are then factorised into the parameters. Thus,
we have
\begin{displaymath}
  \bn_{J*}\sim\mdm(\bn_{J\bullet},\ba)
  \quad\text{and}\quad
  \bn_{J*}\mid\bn_{K*}\sim\mdm(\bn_{J\bullet},\ba+\bnds^{(K)}),
\end{displaymath}
where $\bnds^{(K)} = \{\sum_{i\in K} n_{ia}\}$ for $a=1,\dots,A$,
i.e. the number of $a$ alleles observed for the profiles in $K$.

Finally, when conditioning on the sufficient statistic, $\bnds$, and
the number trails, $\bnsd$, we recover the results for contingency
tables that $n_{**}\mid (\bnds, \bnsd)$ follow a generalisation of the
multivariate-hypergeometric distribution \citep{johnson1997} with
parameters $\bnds$ and $\bnsd$:
\begin{displaymath}
  P(\bn; \bnds, \bnsd) = \dfrac{\prod_{i=1}^I {n_{i\bullet}\choose\bn_i}}%
  {{\ndd\choose \bnds}} =
  \dfrac{\prod_{i=1}^I n_{i\bullet}!\prod_{a=1}^A n_{\bullet a}!}%
  {\ndd!\prod_{i=1}^I\prod_{a=1}^An_{ia}!},
\end{displaymath}
which is identical to Halton's ``exact contingency formula''
\citep{halton1969} and utilised in Patefield's algorithm
\citep{patefield1981} to generate $R\times C$ contingency tables.

\subsection{Moments}
\label{sec:moments}

In Appendix~\ref{app:moments}, we show that the moments of MDM can be
computed using the generalised factorial moments which are given by:
\begin{equation}
  \E\left(\bn^{(\br)}\right) = \mathbb{E}\left\{\prod_{i=1}^I\prod_{a=1}^A
    n_{ia}^{(r_{ia})}\right\} =
  \left\{\prod\limits_{i=1}^I\frac{\nid!}{(\nid{-}\rid)!}\right\}
  \dfrac{\prod\limits_{a=1}^A\prod\limits_{k=0}^{\rda{{-}}1}(\ala{+}k)}%
  {\prod\limits_{k=0}^{\rdd{{-}}1} (\aldd{+}k)},\label{eq:moment}
\end{equation}
where $a^{(b)} = a(a-1)\cdots(a-b+1) = a!/(a-b)!$ is a rising
factorial. Hence, in order to compute the mean of $n_{ia}$, we set
$r_{ia}=1$ and $\rdd=1$ (implying that $\rid=1$ and
$\rda=1$). Plugging this into \eqref{eq:moment}, we obtain
$\mathbb{E}(n_{ia}) = \nid\ala/\aldd = \nid q_a$, as expected.
Furthermore, the covariance matrix can be computed for the different
levels of correlations (left: within individual $i$, and right: between
individuals $i$ and $i'$):
\begin{align*}
  \cov(n_{ia},n_{ia}) &= \nid q_a(1{-}q_a)[1+(\nid{-}1)\theta]
  &\cov(n_{ia},n_{i'a}) &= \nid n_{i'\bullet} q_a(1{-}q_a)\theta\\
  \cov(n_{ia},n_{ia'}) &= -\nid q_aq_{a'}[1+(\nid{-}1)\theta]
  & \cov(n_{ia},n_{i'a'}) &= -\nid n_{i'\bullet} q_aq_{a'}\theta,
\end{align*}
where $\cov(n_{ia},n_{ia})=\var(n_{ia})$. In the case where $\bn_i$
represents a DNA profile, we have for all $i$ that $\nid=2$. Thus, in
this particular case we obtain:
\begin{align*}
  \cov(n_{ia},n_{ia}) &= 2 q_a(1{-}q_a)[1+\theta]
  &\cov(n_{ia},n_{i'a}) &= 4 q_a(1{-}q_a)\theta\\
  \cov(n_{ia},n_{ia'}) &= -2 q_aq_{a'}[1+\theta]
  & \cov(n_{ia},n_{i'a'}) &= -4 q_aq_{a'}\theta,
\end{align*}
which implies positive correlation between counts of identical alleles
within and between individuals. Consequently, for different alleles
the correlation is negative.

\section{Numerical results}
\label{sec:results}

In order to demonstrate how the $\theta$-correction affects
$P(\bm{n}|\mathcal{H})$ in the evaluation of the $L(\mathcal{H})$
expression of Equation 8 in \citet{cowell2014}, we evaluate
\begin{align}
  \text{\sl WoE}(\nda,S_{\bullet a{-}1};Q_a,\theta) &= \frac{P(n_{ia}\mid
    S_{i,a{-}1};Q_a)P(n_{ja}\mid S_{j,a{-}1};Q_a)}%
  {P(n_{ia},n_{ja}\mid S_{i,a{-}1},S_{j,a{-}1};Q_a,\theta)}\notag\\
  &= \frac{Q_a^{\nda}(1{-}Q_a)^{4{-}S_{\bullet a{-}1}{-}n_{\bullet a}}}%
  {\dfrac{\Gamma({\alphadot}_a)}{\Gamma(\alpha_a)\Gamma(\alphadota)}
  \dfrac{\Gamma(\nda{+}\alpha_a)\Gamma(4{-}S_{\bullet a{-}1}{-}n_{\bullet a}{+}\alphadota)}%
  {\Gamma(4{-}S_{\bullet a{-}1}{+}{\alphadot}_a)}}, \label{eq:woe}
\end{align}
where the last expression emphasises that this ratio only depends on
the allele counts $(n_{ia},S_{i,a{-}1})$ and $(n_{ja},S_{j,a{-}1})$
through the \textsl{margins} $\nda=n_{ia}+n_{ja}$ and $S_{\bullet
  a{-}1}=S_{i,a{-}1}+S_{j,a{-}1}$, $0\le \nda+S_{\bullet
  a{-}1}\le4$. Hence, the $\nda=2$ situation covers both the
combination of two heterozygous profiles and also one homozygous
profile together with a profile with no $a$ allele. Similar symmetries
can be identified for different values of $\nda$ and $S_{\bullet
  a{-}1}$.

\begin{figure*}[!h]
  \centering
  \includegraphics[height=\textwidth]{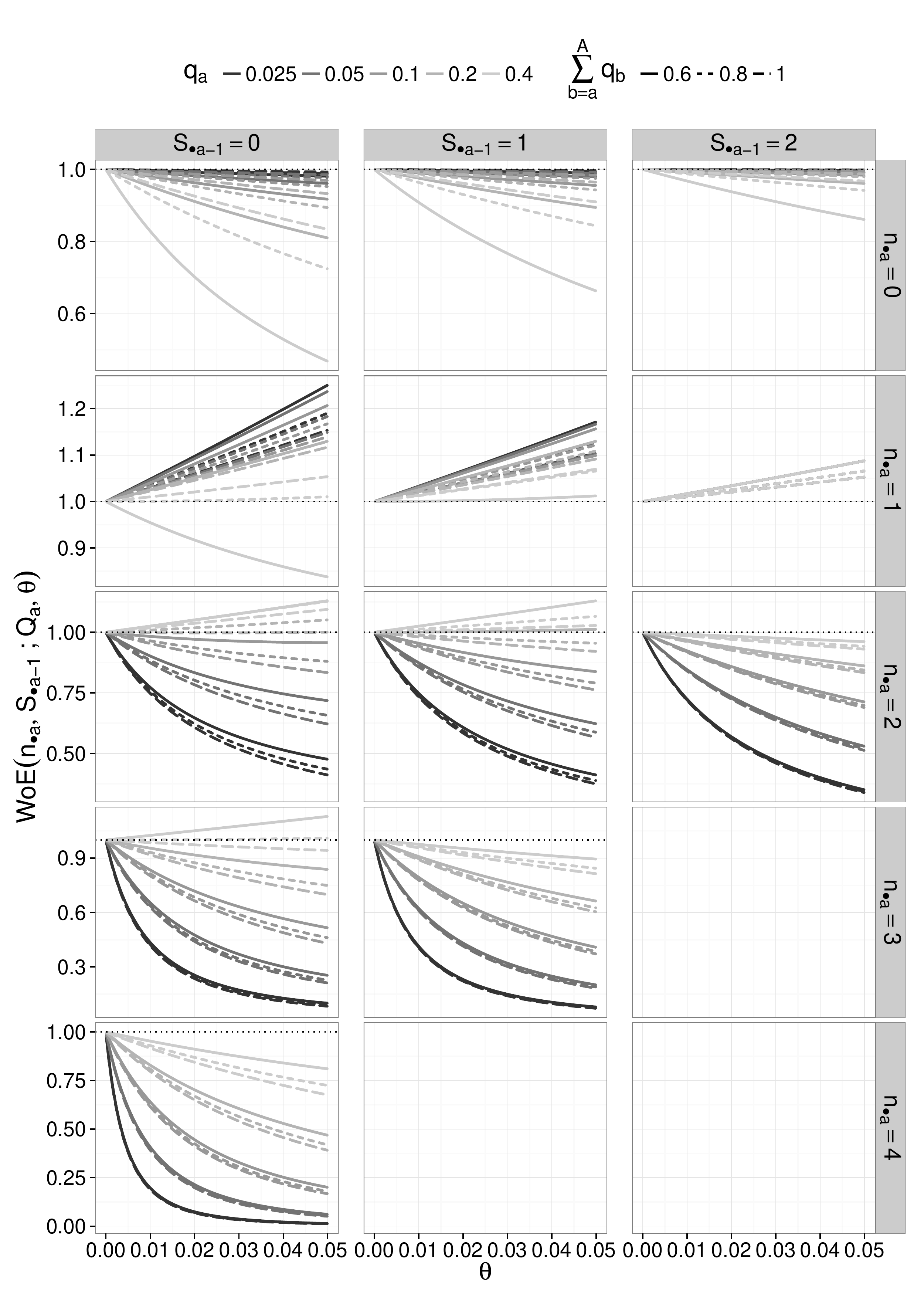}
  \caption{\label{fig:woe2}$\text{\sl WoE}(\nda,S_{\bullet
      a{-}1};Q_a,\theta)$ plotted against $\theta$ for the relevant 12
    combinations for a two-person DNA mixture. The horizontal black
    dotted lines show equal weights of evidences. The variables $\nda$
    and $S_{\bullet a{-}1}$ refer to the margins of the allele counts
    and cumulative allele sums, respectively.}
\end{figure*}

For a two-person DNA mixture only $15$ non-symmetric combinations
exist, although for $S_{\bullet a{-}1}\ge3$ we have that $\nda\le1$,
which implies that no correlation can be observed. Therefore, only
$12$ relevant combinations are shown in Fig.~\ref{fig:woe2}.  The
general picture in Fig.~\ref{fig:woe2} is that\linebreak$\text{\sl
  WoE}(\nda,S_{\bullet a{-}1};Q_a,\theta)<1$, except for $\nda=1$,
where $\text{\sl WoE}(\nda=1,S_{\bullet a{-}1};Q_a,\theta)\ge 1$. That
is, the product of unrelated allele probabilities,
$P(n_{ia}|S_{i,a{-}1})P(n_{ja}|S_{j,a{-}1})$, is smaller than the
joint probability adjusting for relatedness,
$P(n_{ia},n_{ja}|S_{i,a{-}1},S_{j,a{-}1})$. Hence, in the case where
two or more of the same alleles are observed simultaneously, the
weight of evidence is decreased. Conversely, the increased probability
of homozygosity for $\theta>0$ implies that singletons, $\nda=1$, are
less frequent, which implies an increase in the weight of evidence
\citep{buckleton2005}.

\begin{figure*}[!h]
  \centering
  \begin{tabular}{cc}
    \parbox{0.475\textwidth}{\includegraphics[width=0.475\textwidth,page=1]{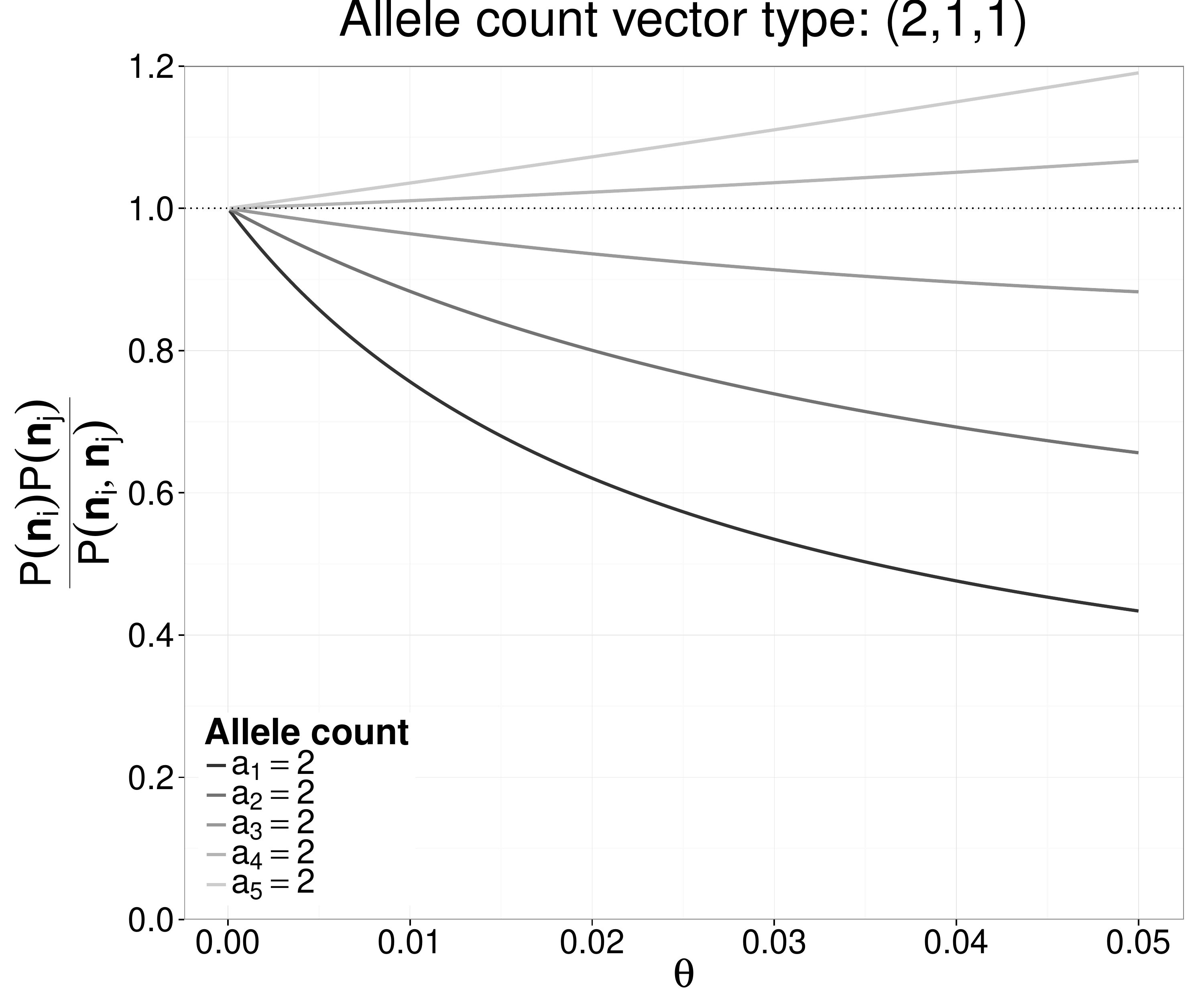}} & 
    \parbox{0.475\textwidth}{\includegraphics[width=0.475\textwidth,page=2]{res2dfRa-bw}} \\
    \addlinespace[5mm]
    \parbox{0.475\textwidth}{\includegraphics[width=0.475\textwidth,page=3]{res2dfRa-bw}} & 
    \parbox{0.475\textwidth}{\includegraphics[width=0.475\textwidth,page=4]{res2dfRa-bw}}    
  \end{tabular}
  \caption{\label{fig:vector}The effect of $\theta$ on the match
    probabilities, $P(\bn_i)P(\bn_j)/P(\bn_i,\bn_j)$, for the possible
    profile combinations for a two-person DNA mixture with shared
    alleles. The underlying allele distribution is
    $\bm{q}=(0.025,0.05,0.1,0.2,0.4)$, where the remaining probability
    mass of $0.225$ is assigned to a ``rest class''.}
\end{figure*}

We also analysed how the ratio between $P(\bm{n}_i)P(\bm{n}_j)$ to
$P(\bm{n}_i,\bm{n}_j)$ behaves. We noted that, due to the
$\theta$-correction, there is an increased probability of shared
alleles among DNA profiles. The behaviour is similar to that pictured
in Fig.~\ref{fig:woe2} since the evaluation is comprised by products
of $\text{\sl WoE}(\nda,S_{\bullet a{-}1};Q_a,\theta)$. In
Fig.~\ref{fig:vector}, we see that it is possible to identify the
contributions from Fig.~\ref{fig:woe2}. For example, the probability
of observing three alleles of one type together with another allele,
$(3,1)$, is the product of $\text{\sl WoE}(\nda,S_{\bullet
  a{-}1};Q_a,\theta)$ for $(\nda,S_{\bullet a{-}1})\in \{(0,0), (1,0)$,
$(3,0), (1,3), (3,1)\}$, which due to the positive correlation between
alleles is dominated by $\text{\sl WoE}(\nda=3,S_{\bullet
  a{-}1}=0;Q_a,\theta)$ and $\text{\sl WoE}(\nda=3,S_{\bullet
  a{-}1}=1;Q_a,\theta)$.

Furthermore, the legend in Fig.~\ref{fig:vector} only specifies the
alleles that were observed more than once because the ratio of
$P(\bm{n}_i)P(\bm{n}_j)$ to $P(\bm{n}_i,\bm{n}_j)$ for alleles
observed only once cancel out. Therefore, the ratio is simplified to a
function of $\theta$, which is independent of the allelic
distribution. For example, in the upper left panel of
Fig.~\ref{fig:vector} (dark grey curve), the ratio is the same for all
vectors $\{(2,1,1,0,0,0), (2,1,0,1,0,0), \dots,$ $(2,0,0,0,1,1)\}$,
i.e.\ all combinations with $a_1$ observed twice give the same ratio.

\section{Conclusion}
\label{sec:conclusion}

We have derived a multivariate generalisation of the
Dirichlet-multinomial distribution for an application in
forensic genetics. The conditional distribution over the cell counts
of the multivariate Dirichlet-multinomial (MDM) distribution also
follows a MDM distribution. Furthermore, the conditional distributions
over vectors follow an extended hypergeometric distribution.

We have demonstrated how to incorporate the $\theta$-correction into
the computational framework of the \texttt{DNAmixtures} package
\citep{dnamixtures2014} and exemplified how the adjustment for positive
correlation between alleles caused by population stratification
affects the weight of evidence.


\appendix 

\section{Generalised factorial moments of MDM}
\label{app:moments}

In this section, we derive the generalised factorial moments of the
MDM distribution. The generalised factorial moments are useful for
count data as it allows relatively simple expressions for most
of the distributions' moments. The generalised factorial moments can
be considered a transformation, $f$, where we use that
$\mathbb{E}\{f(\bn)\} = \sumn f(\bn) P(n)$.

More specifically, $f(\bn) = \prod_{i=1}^I\prod_{a=1}^A
n_{ia}^{(r_{ia})}$, where $a^{(b)} = a!/(a-b)!$ and
$r_{ia}\in\{0,\dots,n_{ia}\}$, is a vector of constants. Hence, if we
want to compute $\mathbb{E}(n_{ia})$, we set $r_{ia}=1$ and all other
to zero.

First, we use that conditioned on $\bm{q}$, the distribution of $\bn$
is found by products of independent multinomial distributions:
\begin{align*}
  \mathbb{E}\left\{\prod_{i=1}^I\prod_{a=1}^A
    n_{ia}^{(r_{ia})}\bigg|\bm{q}\right\} &= 
  \sumn \prod_{i=1}^I\nid! \prod_{a=1}^A
  \frac{n_{ia}^{(r_{ia})}}{n_{ia}!}q_a^{n_{ia}}\\
  &= \sumn \prod_{i=1}^I \frac{\nid!}{(\nid-\rid)!}
  {\nid-\rid\choose\bn_i-\br_i} \prod_{a=1}^A
  q_a^{n_{ia}-r_{ia}}q_a^{r_{ia}}\\
  &= \left\{\prod_{i=1}^I
    \frac{\nid!}{(\nid-\rid)!}\right\}\prod_{a=1}^A q_a^{\rda},
\end{align*}
where we moved terms constant over
$\mathcal{N}=\{\bn:\sum_{a}n_{ia}=\nid\}$ outside the sum and
identified the remaining terms as being the product of independent
multinomial distributions for $\bn-\br$, which by definition sum to unity.

Secondly, we marginalise over $\bm{q}$ in order to obtain the generalised
factorial moments for the multivariate Dirichlet-multinomial
distribution
\begin{align*}
  \mathbb{E}\left\{\prod_{i=1}^I\prod_{a=1}^A
    n_{ia}^{(r_{ia})}\right\} &= 
  \left\{\prod_{i=1}^I \frac{\nid!}{(\nid-\rid)!}\right\}
  \frac{\Gamma(\aldd)}{\prod_{a=1}^A \Gamma(\ala)}
  \int \prod_{a=1}^A q_a^{\rda+\ala-1}\dd \bm{q}\\
  &= \left\{\prod_{i=1}^I \frac{\nid!}{(\nid-\rid)!}\right\}
  \frac{\Gamma(\aldd)}{\Gamma(\aldd+\rdd)} \prod_{a=1}^A
  \frac{\Gamma(\ala+\rda)}{\Gamma(\ala)}.
\end{align*}
For the remaining terms, we see that the ratios of gamma functions
involve $\Gamma(\beta+t)$ and $\Gamma(\beta)$. For $t>0$, the gamma
function satisfies
\begin{displaymath}
  \frac{\Gamma(\beta+t)}{\Gamma(\beta)} = \prod_{k=0}^{t-1}(\beta+k).
\end{displaymath}
Hence, the expression for $\E(\bn^{(\br)})$ may be simplified to
\begin{displaymath} 
  \E\left(\bn^{(\br)}\right) = \mathbb{E}\left\{\prod_{i=1}^I\prod_{a=1}^A
    n_{ia}^{(r_{ia})}\right\} =
  \left\{\prod\limits_{i=1}^I\frac{\nid!}{(\nid-\rid)!}\right\}
  \dfrac{\prod\limits_{a=1}^A\prod\limits_{k=0}^{\rda{-}1}(\ala+k)}%
  {\prod\limits_{k=0}^{\rdd{-}1} (\aldd+k)}.
\end{displaymath}

\bibliographystyle{chicago}
\bibliography{cite}
\end{document}